\documentclass{IEEEtran}
\usepackage{amsmath}
\usepackage{amssymb}  
\usepackage{graphicx}
\usepackage{listings}
\usepackage[utf8]{inputenc}
\usepackage{algorithm}
\usepackage{algpseudocode}
\usepackage{xcolor}
\usepackage{siunitx}
\usepackage{float}
\usepackage[hidelinks]{hyperref}
\usepackage{url}
\usepackage{verbatim}
\usepackage{xspace}   
\usepackage{array}    
\title{Semicustom Frontend VLSI Design and Analysis of a 32-bit Brent-Kung Adder in Cadence Suite}

\author{\IEEEauthorblockN{\textbf{Yashvardhan Singh }}\\
\IEEEauthorblockA{Department of Electronics and Communication Engineering, MIT, Manipal \\
Email: yvs.373@gmail.com}}

\begin{document}
\maketitle

\begin{abstract}
Adders are fundamental components in digital circuits, playing a crucial role in arithmetic operations within computing systems and many other applications. This paper focuses on the design and simulation of a 32-bit Brent-Kung parallel prefix adder, which is recognized for its efficient carry propagation and logarithmic delay characteristics. The Brent-Kung architecture balances computational speed and hardware complexity, making it suitable for high-speed digital applications. The design is implemented using Verilog HDL and simulated using Cadence Design Suite tools, including NCLaunch and Genus, to evaluate its performance in terms of scalability, speed, and functional working. Comparative analysis with traditional adder architectures highlights the advantages of the Brent-Kung adder for modern digital systems.
\\ Keywords: Brent-Kung Adder; Parallel Prefix Adder; Verilog; Digital Circuits; Semicustom VLSI Design; Cadence Design Suite
\end{abstract}

\section{Introduction}
Adders are logic circuits serving as building blocks for arithmetic operations in digital computing systems. They are combinational logic circuits that perform binary addition, producing a sum and a carry output based on the inputs provided. Over time, various types of adders have been developed to address the growing demands for speed, efficiency, and scalability in digital systems. Several prominent adder architectures are discussed below.

\subsection{Half Adder}
The half adder is the simplest form of an adder, capable of adding two single-bit binary numbers. It produces two outputs: the sum and the carry. While it is a critical building block for more complex adders, its limitation lies in its inability to account for carry inputs from previous stages.\begin{figure}[H]
    \centering
    \includegraphics[width=0.5\linewidth]{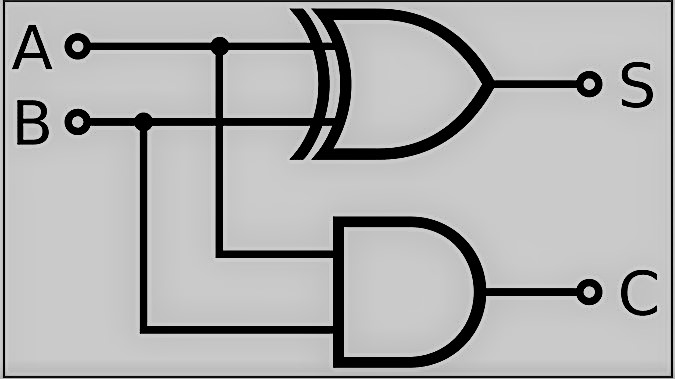}
    \caption{Half-Adder Logic Gate Implementation }
    \label{fig:ha-diagram}
\end{figure}

\subsection{Full Adder}
To overcome the limitations of the half adder, the full adder was introduced. It can add three 1-bit binary numbers, two operands and a carry input, producing a sum and a carry output. Full adders are often cascaded to form multi-bit adders for larger binary numbers.
\begin{figure}[H]
    \centering
    \includegraphics[width=0.75\linewidth]{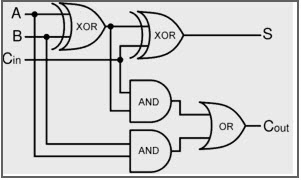}
    \caption{Full-Adder Logic Gate Implementation }
    \label{fig:Full Adder Logic Diagram}
\end{figure}

\subsection{Ripple Carry Adders}
When multiple full adders are connected in parallel, they form a binary parallel adder capable of adding multi-bit binary numbers simultaneously. However, this design introduces a significant limitation: carry propagation delay. In ripple-carry adders, each stage must wait for the carry from the previous stage, resulting in slower performance as the number of bits increases.
\begin{figure}[H]
    \centering
    \includegraphics[width=0.85\linewidth]{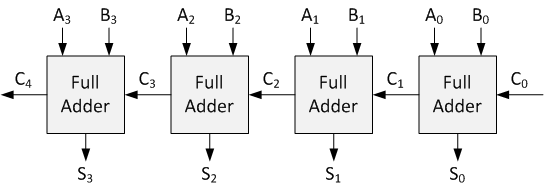}
    \caption{Ripple Carry Adder}
    \label{fig:Ripple  Carry Adder Logic Diagram}
\end{figure}

\subsection{Carry Look Ahead Adders}
To address the delay caused by ripple-carry propagation, the carry lookahead adder was developed. This design uses additional logic to compute carries in advance, significantly reducing delay and improving speed. Despite its advantages, carry lookahead adders require more complex hardware, which can become inefficient for very large bit-widths.
\begin{figure}[H]
    \centering
    \includegraphics[width=0.85\linewidth]{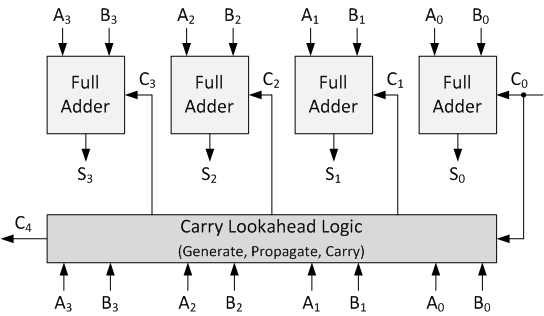}
    \caption{Carry Look Ahead Adder}
    \label{fig:Carry Look Ahead Adder}
\end{figure}
\subsection{The Need for faster Adders}
As digital systems evolved, so did the demand for faster and more efficient arithmetic operations. Applications such as high-speed processors, signal processing units, and real-time systems required adders with minimal delay and optimized resource usage. Although carry look-forward adders improved performance significantly, their complexity and fan-out issues posed challenges for scalability.

\subsection{Brent-Kung Adder}
The 32-bit Brent-Kung Adder is a parallel prefix adder designed with a logarithmic-depth tree structure, enabling faster carry computation compared to ripple-carry adders. This design reduces computational delay and fan-out compared to traditional adders like ripple-carry or carry lookahead adders. By balancing speed and hardware complexity, Brent-Kung adders are well-suited for high-speed applications requiring efficient arithmetic operations.

 This paper focuses on designing and simulating a 32-bit Brent-Kung adder using Verilog HDL and Cadence NCLaunch to demonstrate its advantages in terms of speed, scalability, and efficiency.
\begin{figure}[H]
    \centering
    \includegraphics[width=1\linewidth]{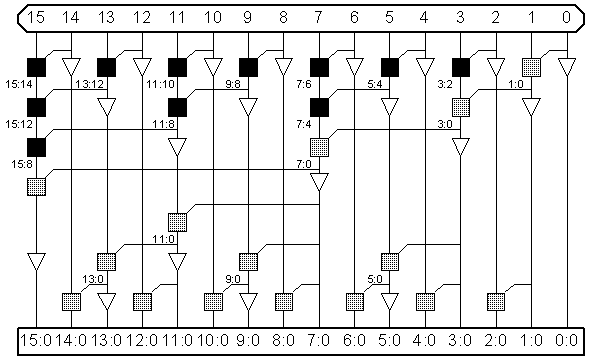}
    \caption{Brent Kung Adder (16 bit)}
    \label{fig:Brent Kung Adder (16 bit)}
\end{figure}

\section{Brief Description - Brent-Kung Adder}
The 32-bit Brent-Kung Adder is a logarithmic-depth parallel prefix adder designed to compute carries faster than ripple-carry adders due to its tree-like structure.

Designed by Richard P. Brent and H.T. Kung in 1982 [1], the Brent-Kung Adder (BKA) is a well-known parallel prefix adder that provides an optimal number of stages from input to all outputs while minimizing wiring complexity. 

BKA occupies less area than other adders, such as the Sparse Kogge Stone Adder (SKA), Kogge-Stone adder (KSA), and Spanning tree adder. The BKA also uses a limited number of propagating and generating cells, further contributing to its efficiency [13].

\subsection{Stages of Brent-Kung Adders}
\begin{itemize}
    \item \textbf{Pre-processing Stage:} Computes Generate (G) and Propagate (P) signals:\begin{equation}
        G_i = A_i \cdot B_i, \quad P_i = A_i \oplus B_i
    \end{equation}
      \item \textbf{Prefix Carry Tree Stage:} Uses Black and Gray cells to compute carry signals:\begin{equation}
        G_{i:j} = G_{i:k} + P_{i:k} G_{k-1:j}, \quad P_{i:j} = P_{i:k} P_{k-1:j}
    \end{equation}
     \item \textbf{Post-processing Stage:} Generates sum bits using computed carry signals.
\end{itemize}

 \begin{figure}[H]
            \centering
\includegraphics[width=0.85\linewidth]{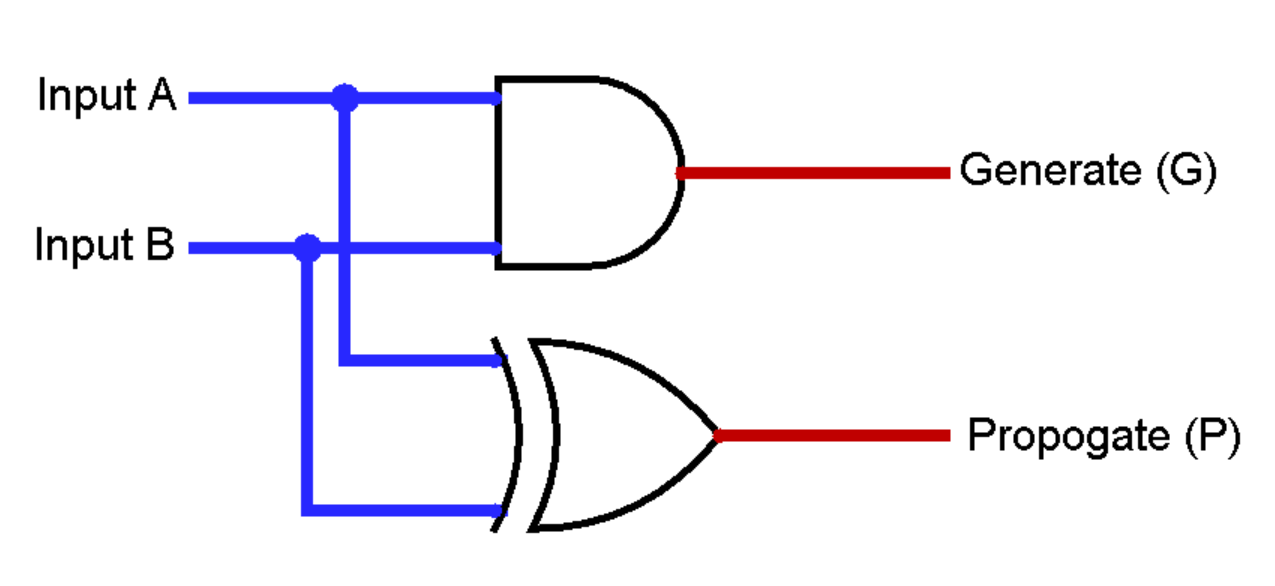}
    \caption{Pre Processing Logic Circuit}
    \label{fig:Pre Processing Logic Circuit}
\end{figure}

    \begin{figure}[H]
            \centering
            \includegraphics[width=0.85\linewidth]{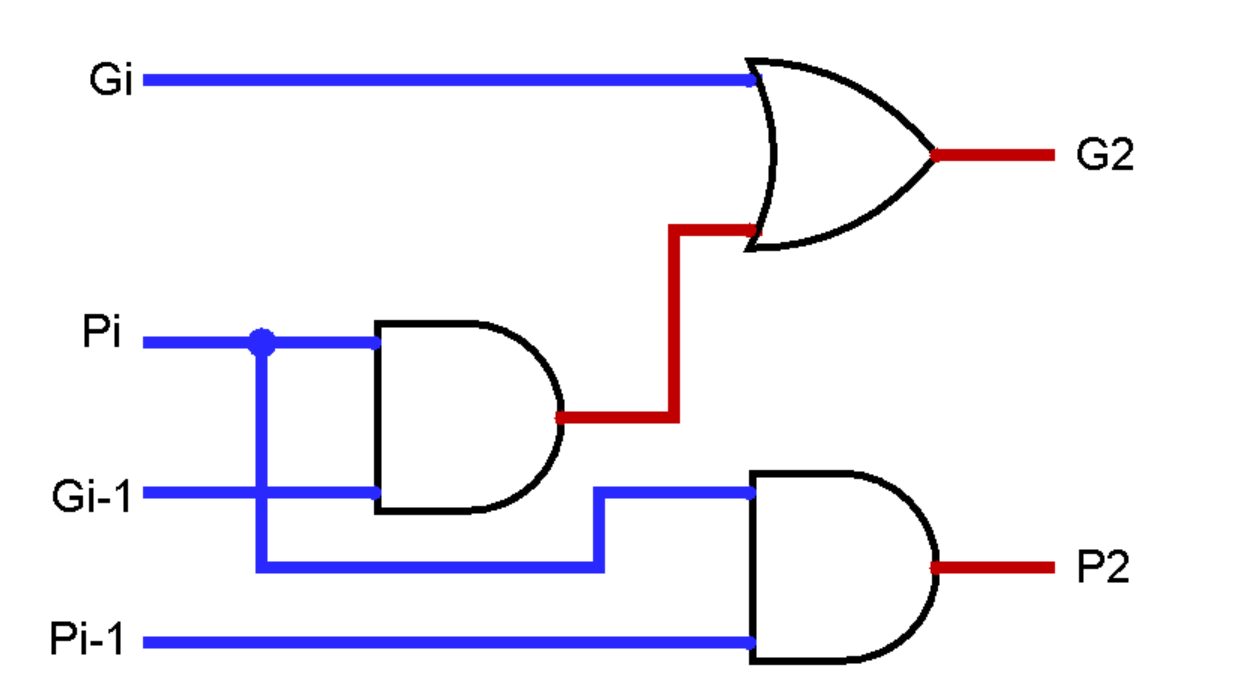}
            \caption{Black Cell}
            \label{fig:Black Cell}
        \end{figure}

\begin{figure}[H]
    \centering
\includegraphics[width=0.85\linewidth]{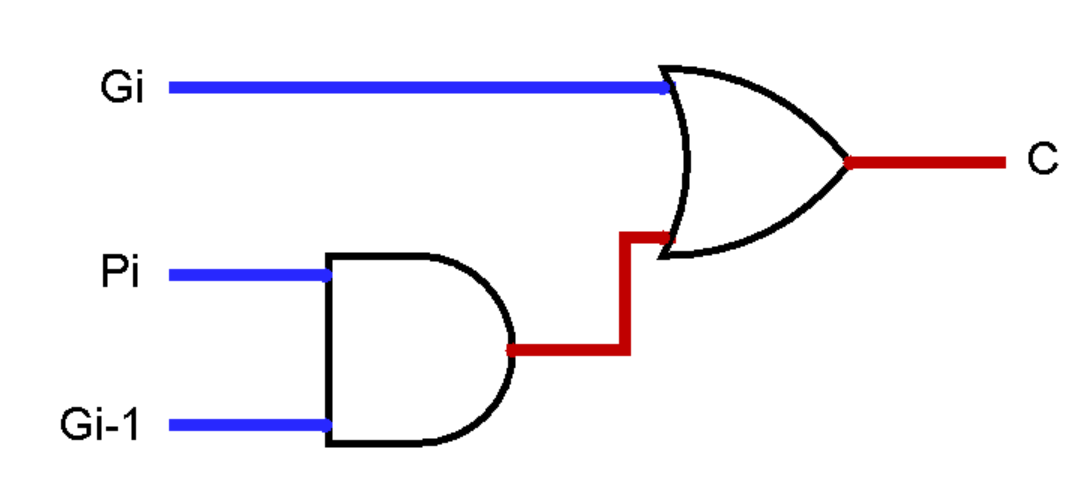}
    \caption{Gray Cell}
    \label{fig:Gray Cell}
\end{figure}

\begin{figure}[H]
    \centering
\includegraphics[width=0.6\linewidth]{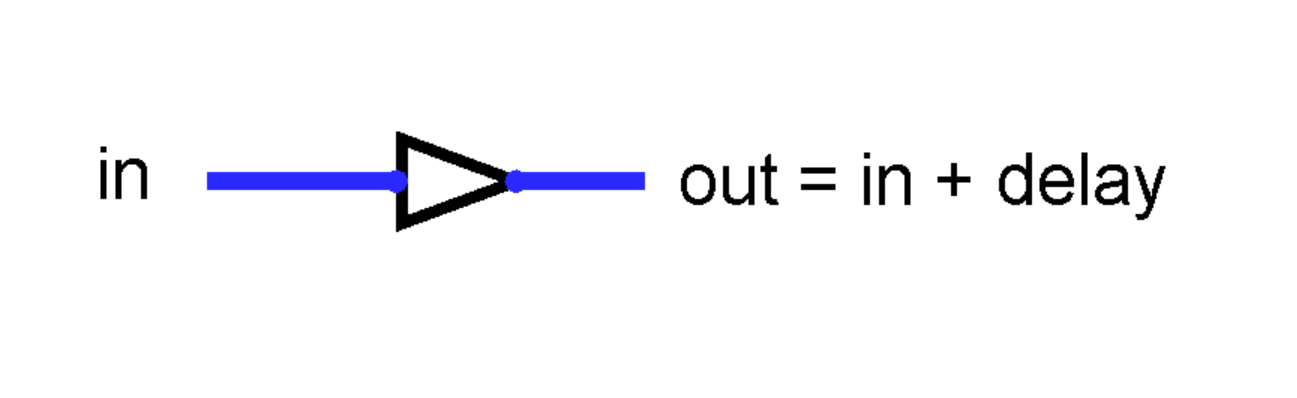}
    \caption{White Cell}
    \label{fig:White Cell}
\end{figure}

\begin{figure}[H]
    \centering
\includegraphics[width=0.7\linewidth]{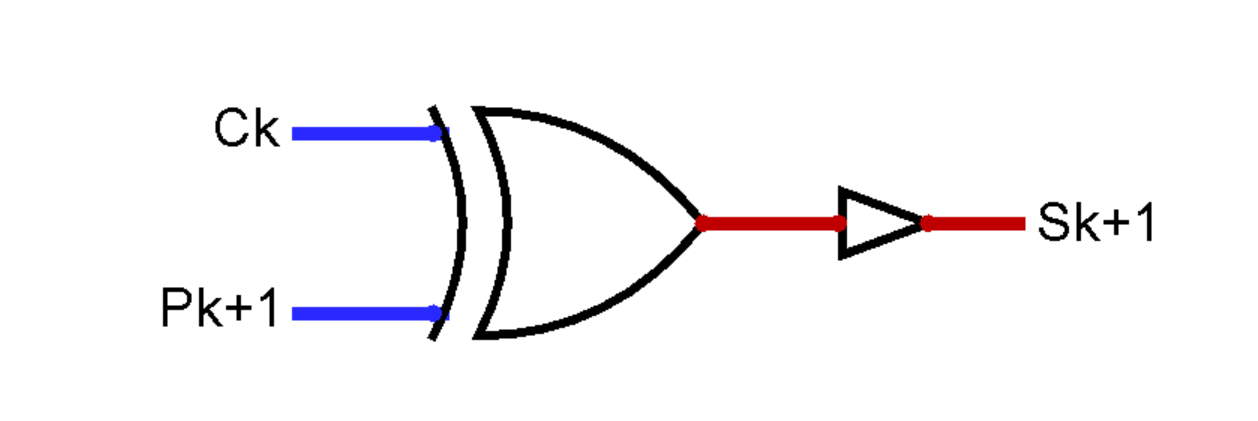}
    \caption{Post Processing Cell}
    \label{fig:Post Processing Cell}
\end{figure}

\vspace{-0.5cm}

In the pre-processing stage, each pair of input bits $A_i$ and $B_i$ is processed to generate two signals: the Generate (G) and Propagate (P) signals. 

The Carry Prefix Tree, an intermediate stage, is crucial as it efficiently computes carry signals using a structured arrangement of logic cells called Black cells and Gray cells. The data flow in this stage follows a tree-like hierarchical structure. The propagate and generate signals from the pre-processing stage serve as inputs to these cells. 

This hierarchical arrangement significantly reduces carry propagation delay by computing carries in parallel rather than sequentially. 

In the post-processing stage, the sum bits are computed by combining the propagate signals from the pre-processing stage with carry outputs from the prefix carry tree stage. The flow of data through the adder is aptly visualized in the flowchart shown in Fig. 11.
\\
\begin{figure}[H]
    \centering
    \includegraphics[width=0.6\linewidth]{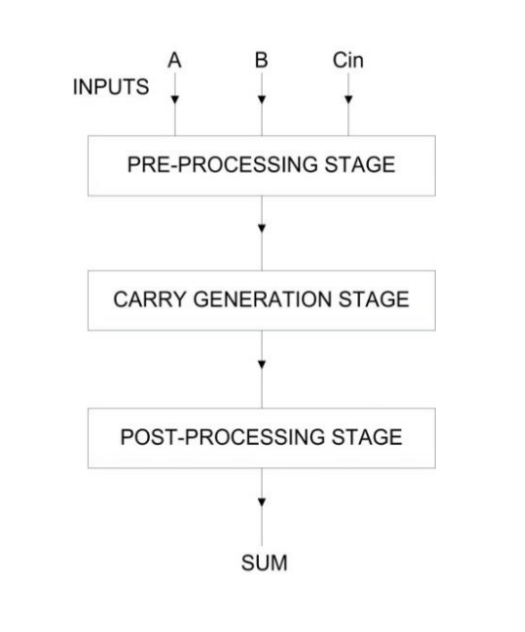}
    \caption{Flow of data within Brent Kung Adder}
    \label{fig:Flowdata}
\end{figure}

\subsection{Comparing the Brent-Kung adder with other adders}
\begin{itemize}
\item \textbf{Ripple-Carry Adder:} High delay due to sequential carry propagation.
\item \textbf{Carry-Lookahead Adder:} Faster but more complex in terms of hardware.
\item \textbf{Brent-Kung Adder:} Faster than RCA and CLA, and balances speed and hardware complexity, making it ideal for high-speed applications.
\item \textit{To gain a deeper understanding of the gate-level operations and verify design principles, a small-scale 4-bit Brent-Kung adder was implemented using Logisim Evolution software. This preliminary design served as a valuable learning tool and provided a clear visualization of the adder's logic structure before scaling up to the full 32-bit RTL implementation.}
\begin{figure}[H]
    \centering
    \includegraphics[width=1\linewidth]{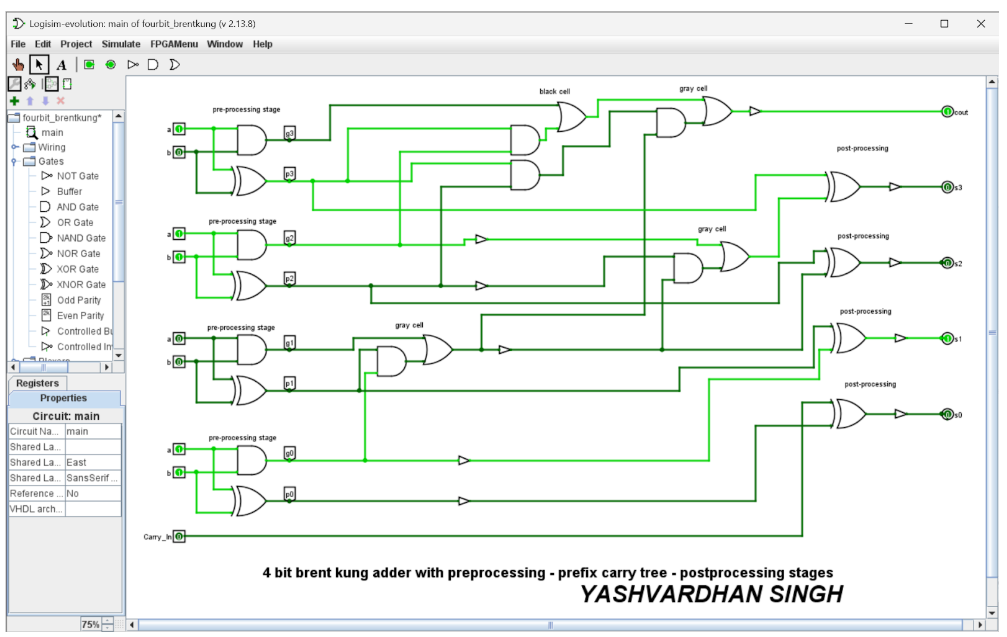}
    \caption{Functional Analysis of 4-bit BKA in Logisim}
    \label{fig:enter-label}
\end{figure}
\end{itemize}
\vspace{-0.6cm}

\section{Circuit Design}

\subsection{Schematic Design}
The BKA consists of three main sub-circuits: Black cell, Gray cell, and White Cell(Buffer), as discussed previously. These are modeled separately as modules in Verilog HDL and integrated into the main/top BKA module. A schematic view for the same can be seen in Fig. 13. 
\begin{figure}[H]
    \centering
    \includegraphics[width=1\linewidth]{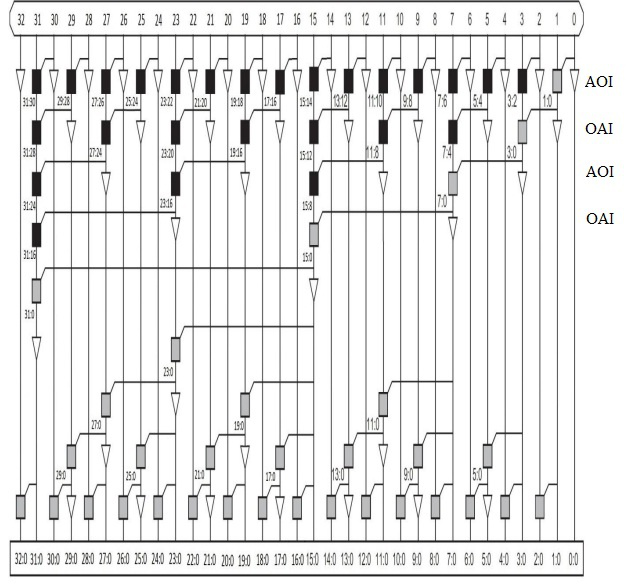}
    \caption{32 Bit Brent-Kung Adder Schematic}
    \label{fig:enter-label}
\end{figure}
\subsection{Verilog Implementation}
The Brent-Kung adder was implemented in Verilog HDL with a modular approach, and Algorithm 1 explains the execution.
\\
\begin{algorithm}
\caption{Brent-Kung Adder (32-bit) - Pseudocode}
\label{algo:bka}

\textbf{MODULE: Black Cell} \\
\textbf{INPUT:} $g_{ik}, p_{ik}, g_{kj}, p_{kj}$ \\
\textbf{OUTPUT:} $g_{ij}, p_{ij}$ \\
\textbf{PROCESS:}
\begin{enumerate}
    \item Compute carry signal: $g_{ij} = g_{ik} \lor (p_{ik} \land g_{kj})$
    \item Compute propagate signal: $p_{ij} = p_{ik} \land p_{kj}$
\end{enumerate}

\vspace{0.5em}

\textbf{MODULE: Gray Cell} \\
\textbf{INPUT:} $g_i, p_i, g_{i-1}$ \\
\textbf{OUTPUT:} $c$ \\
\textbf{PROCESS:}
\begin{enumerate}
    \item Compute carry signal: $c = g_i \lor (p_i \land g_{i-1})$
\end{enumerate}

\vspace{0.5em}

\textbf{MODULE: White Cell} \\
\textbf{INPUT:} $a$ \\
\textbf{OUTPUT:} $y$ \\
\textbf{PROCESS:}
\begin{enumerate}
    \item Assign value: $y = a$
\end{enumerate}

\vspace{0.5em}

\textbf{MODULE: Preprocessing} \\
\textbf{INPUT:} $a, b$ \\
\textbf{OUTPUT:} $g, p$ \\
\textbf{PROCESS:}
\begin{enumerate}
    \item Compute generate signal: $g = a \land b$
    \item Compute propagate signal: $p = a \oplus b$
\end{enumerate}

\vspace{0.5em}

\textbf{MODULE: Postprocessing} \\
\textbf{INPUT:} $c, p$ \\
\textbf{OUTPUT:} $s$ \\
\textbf{PROCESS:}
\begin{enumerate}
    \item Compute sum bit: $s = c \oplus p$
\end{enumerate}

\vspace{0.5em}

\textbf{MODULE: Brent-Kung Adder (32-bit)} \\
\textbf{INPUTS:} $A[31:0], B[31:0], C_i$ (Carry-in) \\
\textbf{OUTPUTS:} $S[31:0]$ (Sum), $C_o$ (Carry-out) \\

1. Preprocessing Stage:
   \begin{itemize}
       \item FOR each bit $i$ from 0 to 31:
       \begin{itemize}
           \item Compute propagate ($P_1[i]$) and generate ($G_1[i]$) signals using Preprocessing module.
       \end{itemize}
   \end{itemize}

2. Prefix Tree for Carry Computation:
   \begin{itemize}
       \item FOR each stage from 2-bit to 32-bit groups:
       \begin{itemize}
           \item Use Black Cell modules to compute higher-order propagate and generate signals.
       \end{itemize}
       Final Stage:
       \begin{itemize}
           \item Use Black Cell module to compute final carry signals.
       \end{itemize}
   \end{itemize}

3. Carry Computation:
   Initialize $C[0] = C_i$ (Carry-in).\\
   FOR each bit $i$ from 1 to 32:
   Use Gray Cell modules to compute intermediate carry signals.

4. Postprocessing Stage:
   FOR each bit $i$ from 0 to 31:
   Compute final sum using Postprocessing module.

Assign $C_o = C[32]$ (Final Carry-out).

END MODULE

\end{algorithm}
The complete implementation, including the main adder module and testbench, is available in our GitHub repository.
\vspace{-0.5cm}

\subsection{Verilog Implementation - Testbench}

The adder was verified using a testbench, comprising of six comprehensive test cases:
\vspace{0.2cm}
\begin{itemize}
    \item \textbf{Basic Addition:} Adding $1+1$ to verify fundamental operation.
    \item \textbf{Overflow Handling:} Testing maximum 32-bit value ($4,294,967,295 + 1$).
    \item \textbf{Carry Propagation:} Adding two large numbers ($2,147,483,648 + 2,147,483,648$).
    \item \textbf{Zero Addition:} Verifying the $0+0$ edge case.
    \item \textbf{Random Values:} Testing arbitrary values ($10+20$).
    \item \textbf{Carry-in Effect:} Examining carry-in influence ($15+1+\text{Cin}$).
\end{itemize}

All test cases were simulated and thoroughly analyzed in subsequent sections.
\vspace{-0.5cm}

\section{Simulation and Synthesis}
\vspace{-0.91cm}
\subsection{ Simulation}
Initial testing and verification were carried out on EDAplayground using IcarusVerilog 12.0 with -Wall -g2012. Outputs were observed via \&monitor logs and EPWave waveforms. Verification was then performed on the Cadence
NCLaunch software, which gave the same results as the Icarus
Verilog simulation. This adds to the accuracy of our previous
results, which is discussed in great detail in subsequent sections.

\begin{figure}[H]
    \centering
    \includegraphics[width=1\linewidth]{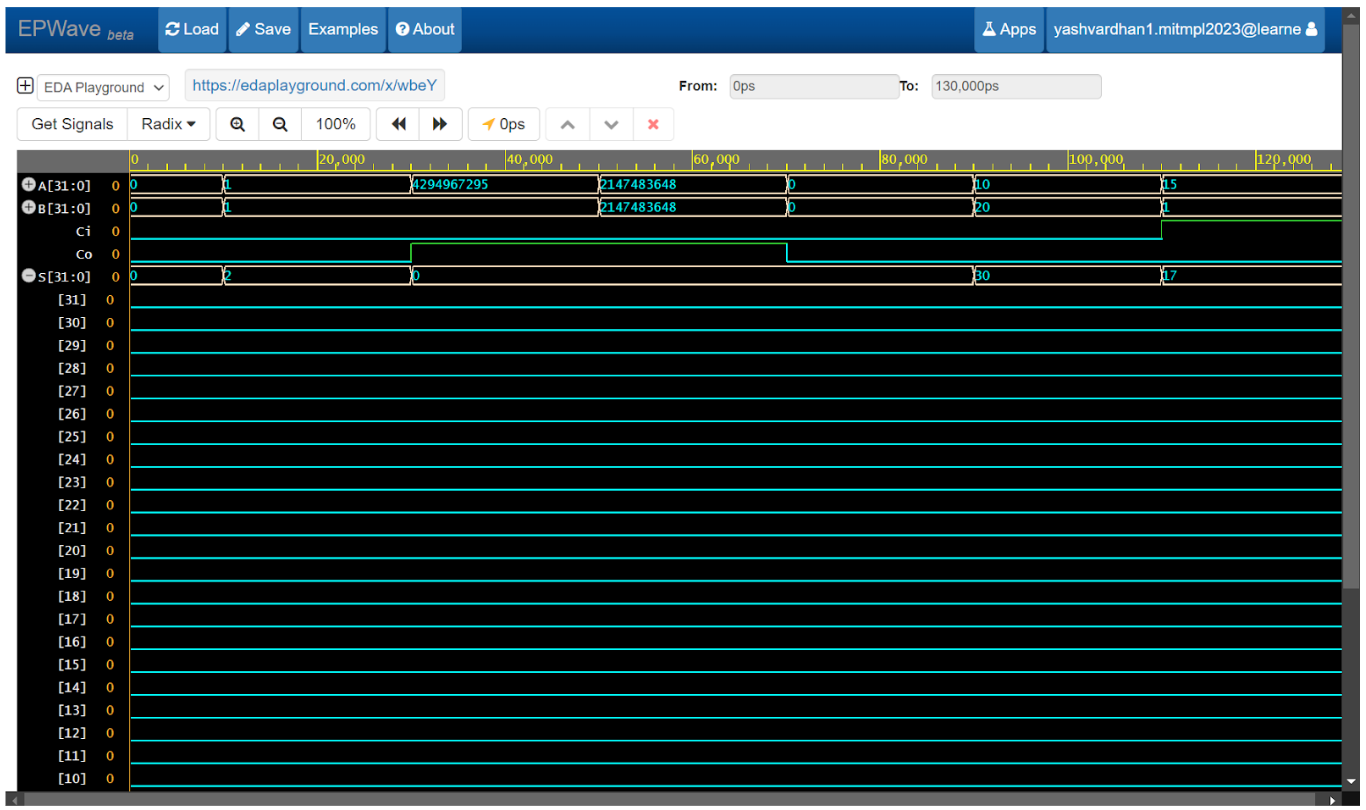}
    \caption{Icarus Verilog - EDA-Playground Simulation - EPWave}
    \label{fig:enter-label}
\end{figure}
\begin{figure}[H]
    \centering
    \includegraphics[width=1\linewidth]{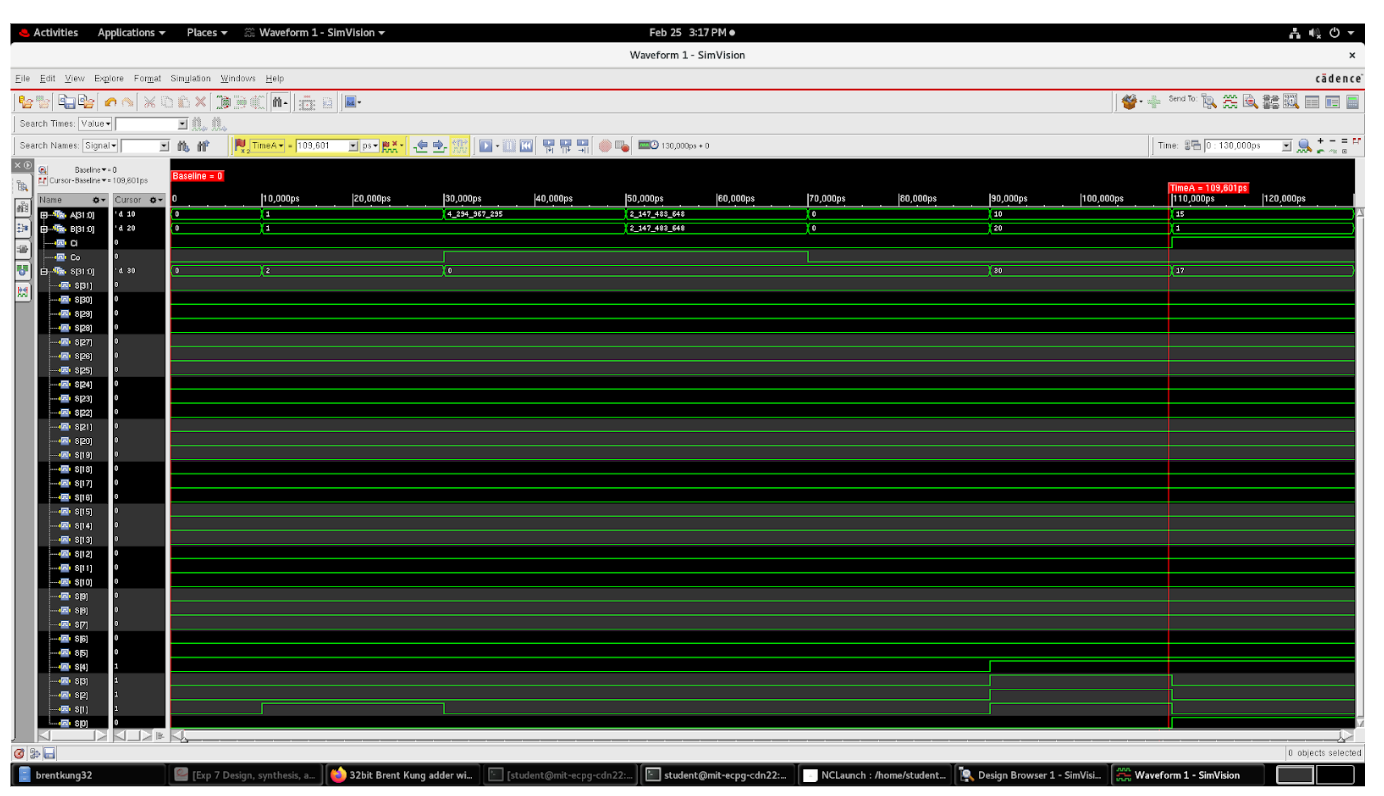}
    \caption{Cadence NC-Launch Simulation}
    \label{fig:enter-label}
\end{figure}
\vspace{-0.6cm}
\subsection{Initial Simulation: Detailed Analysis}

\noindent The simulation results in Table 1 demonstrate the 32-bit Brent-Kung adder's functionality across several test cases. Test 1 verified the initial state with zero inputs, producing the expected zero output. Test 2 confirms basic addition functionality by correctly computing 1+1=2. Tests 3 and 4 validated overflow handling: adding the maximum 32-bit value (4.29B)(B indicates Billion) to 1, and adding two large numbers (2.14B+2.14B), both correctly produced zero sum with carry-out=1, demonstrating proper overflow detection. Test 5 rechecked the zero-addition edge case. Test 6 verified the correct operation with arbitrary values (10+20=30). Finally, Test 7 confirmed the adder's ability to process carry-in signals by correctly computing 15+1+1=17.  All test cases demonstrate that the Brent-Kung adder is functioning correctly, handling various scenarios. \\
This completes the Functional Analysis of our design. In the following section, let us explore performance in terms of parameters like power, area, and time.
\begin{table}[h]
    \centering
    \small
    \renewcommand{\arraystretch}{1.1}
    \setlength{\tabcolsep}{2pt} 
    \begin{tabular}{|c|r|r|c|r|c|r|}
        \hline
        \textbf{Test} & \textbf{A} & \textbf{B} & \textbf{Cin} & \textbf{Sum} & \textbf{Cout} & \textbf{Time (ns)} \\ \hline
        1 & 0 & 0 & 0 & 0 & 0 & 0 \\ \hline
        2 & 1 & 1 & 0 & 2 & 0 & 10 \\ \hline
        3 & 4.29B & 1 & 0 & 0 & 1 & 30 \\ \hline
        4 & 2.14B & 2.14B & 0 & 0 & 1 & 50 \\ \hline
        5 & 0 & 0 & 0 & 0 & 0 & 70 \\ \hline
        6 & 10 & 20 & 0 & 30 & 0 & 90 \\ \hline
        7 & 15 & 1 & 1 & 17 & 0 & 110 \\ \hline
    \end{tabular}
    \caption{Simulation Results for 32-bit Brent-Kung Adder}
    \label{tab:simulation_results}
\end{table}

\subsection{Synthesis using Cadence Genus}
Power, Area, and Timing (PAT) analysis was performed using Cadence Genus 21.14 after synthesizing the design. The Verilog code from previous sections was synthesized using Cadence Genus 21.14 with a 90nm standard-cell library, specifically using the slow corner library for conservative performance estimates.
\\
Due to the combinational nature of the design, unconstrained timing conditions were applied via an TCL file to enable accurate timing analysis. No explicit timing constraints (.sdc file) were specified during synthesis. Instead, the -unconstrained option in Genus was used, allowing the tool to perform default timing optimization without predefined frequency or timing targets.

\begin{figure}[H]
    \centering
    \includegraphics[width=1\linewidth]{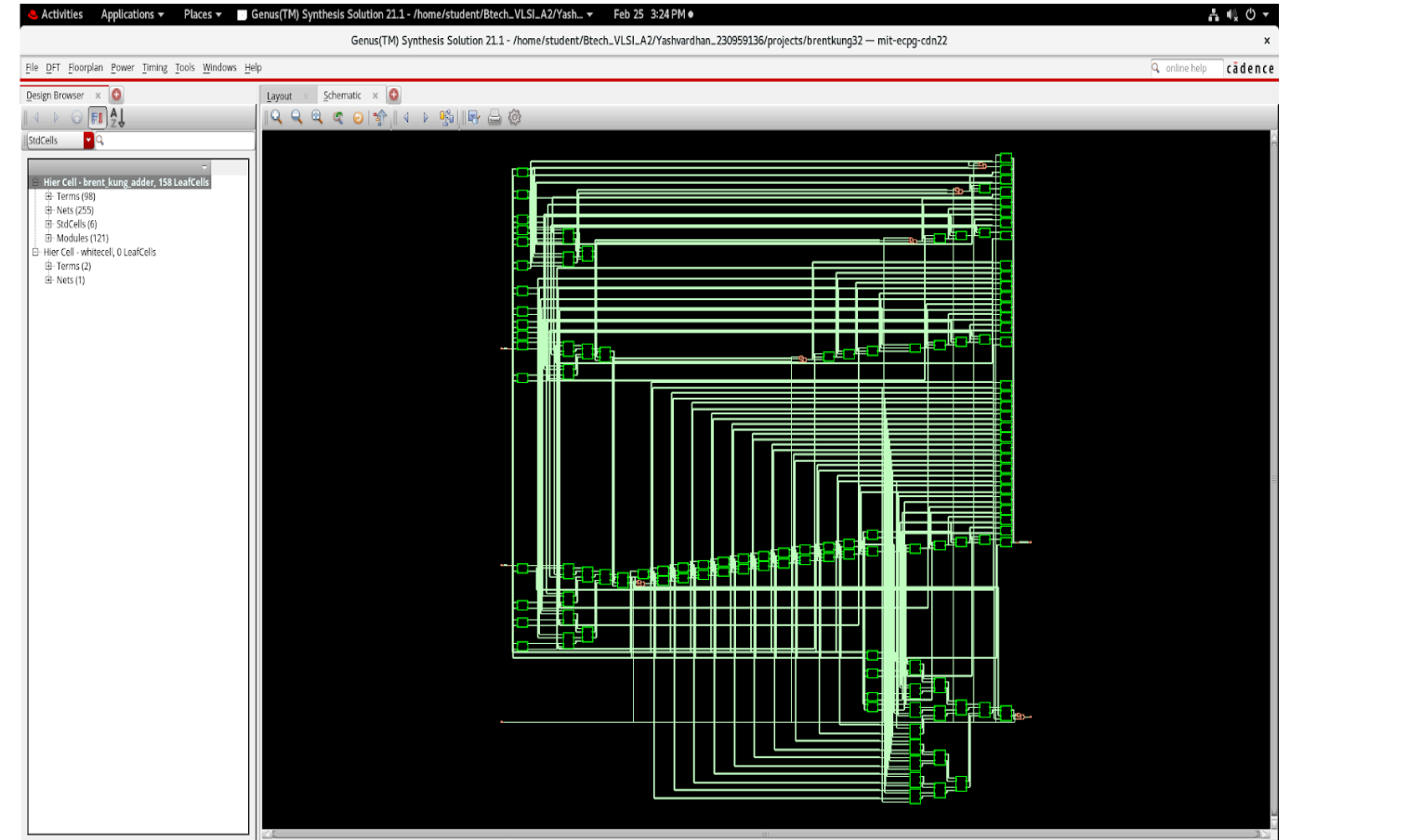}
    \caption{Brent Kung Adder Design Synthesis using Cadence Genus}
    \label{fig:enter-label}
\end{figure}
\vspace{-0.3cm}
\subsubsection{\textbf{Power Report}}
The synthesized design consumes a total power of \SI{43.32}{\micro\watt}, with the following power components:  
\begin{itemize}
    \item \textbf{Leakage Power:} \SI{8.63}{\micro\watt} (\SI{19.93}{\percent})
    \item \textbf{Internal Power:} \SI{26.03}{\micro\watt} (\SI{60.09}{\percent})
    \item \textbf{Switching Power:} \SI{8.66}{\micro\watt} (\SI{19.98}{\percent})
\end{itemize}
This indicates that internal power dominates total consumption, followed by switching and leakage power.
The obtained Power Report is as follows:
\begin{figure}[H]
    \centering
    \includegraphics[width=0.6\linewidth]{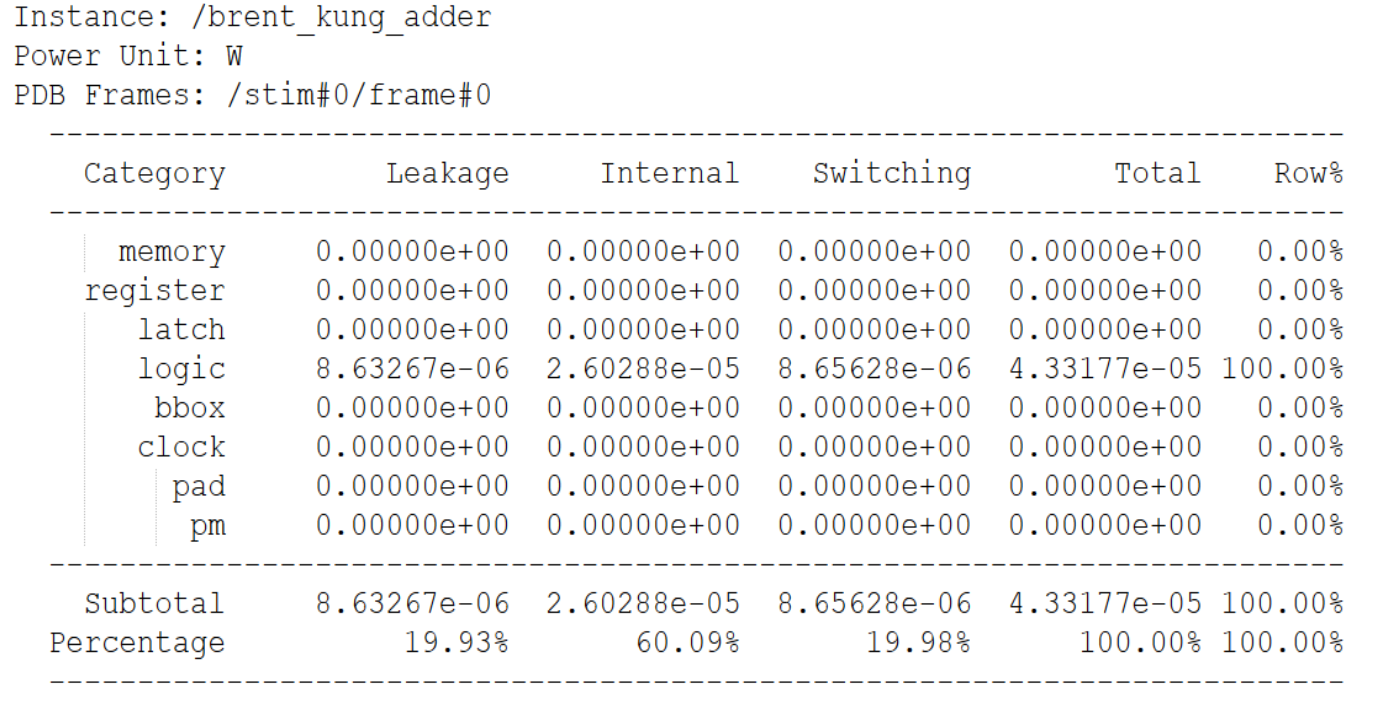}
    \caption{Power Report}
    \label{fig:enter-label}
\end{figure}
\subsubsection{\textbf{Area Report}}
Total cell area of 1223.91 $\mu$m².
The obtained Area Report is as follows:
\begin{figure}[H]
    \centering
    \includegraphics[width=1\linewidth]{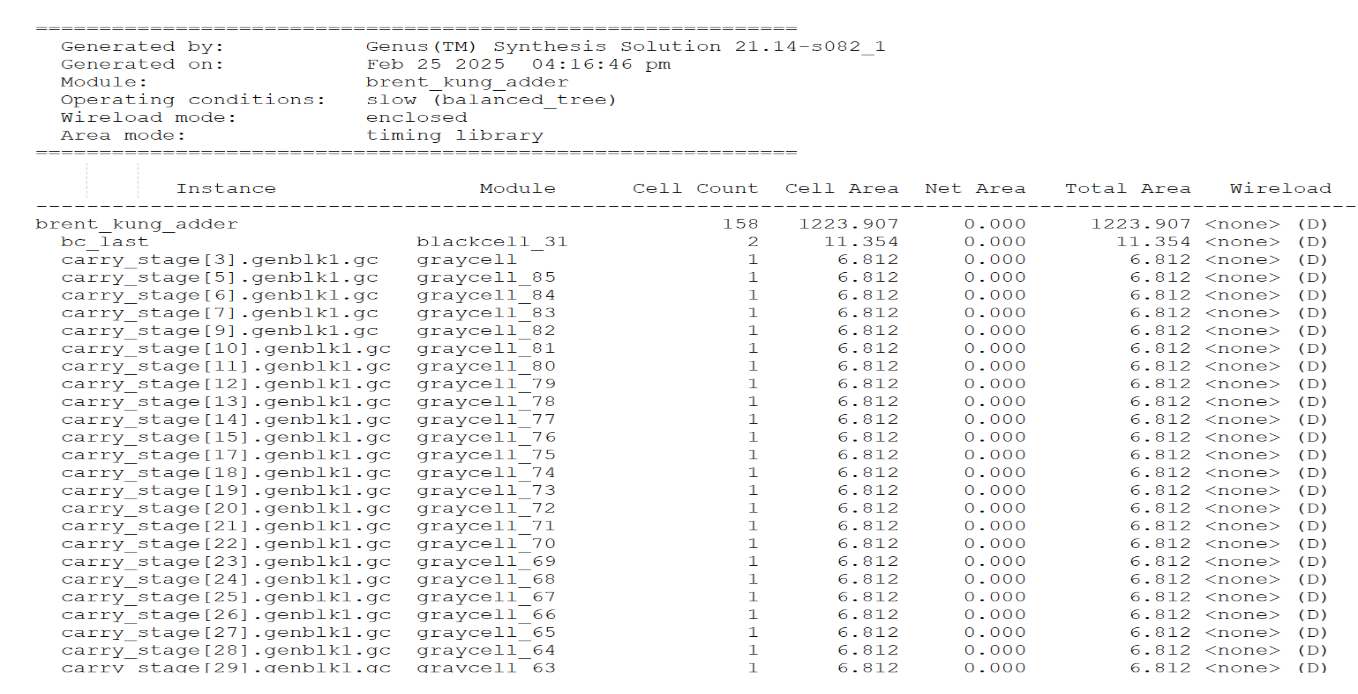}
    \caption{Area Report}
    \label{fig:enter-label}
\end{figure}
\subsubsection{\textbf{Timing Report}}
The critical path delay is of 3.78 ns.
The obtained Timing Report is as follows:
\begin{figure}[H]
    \centering
    \includegraphics[width=1\linewidth]{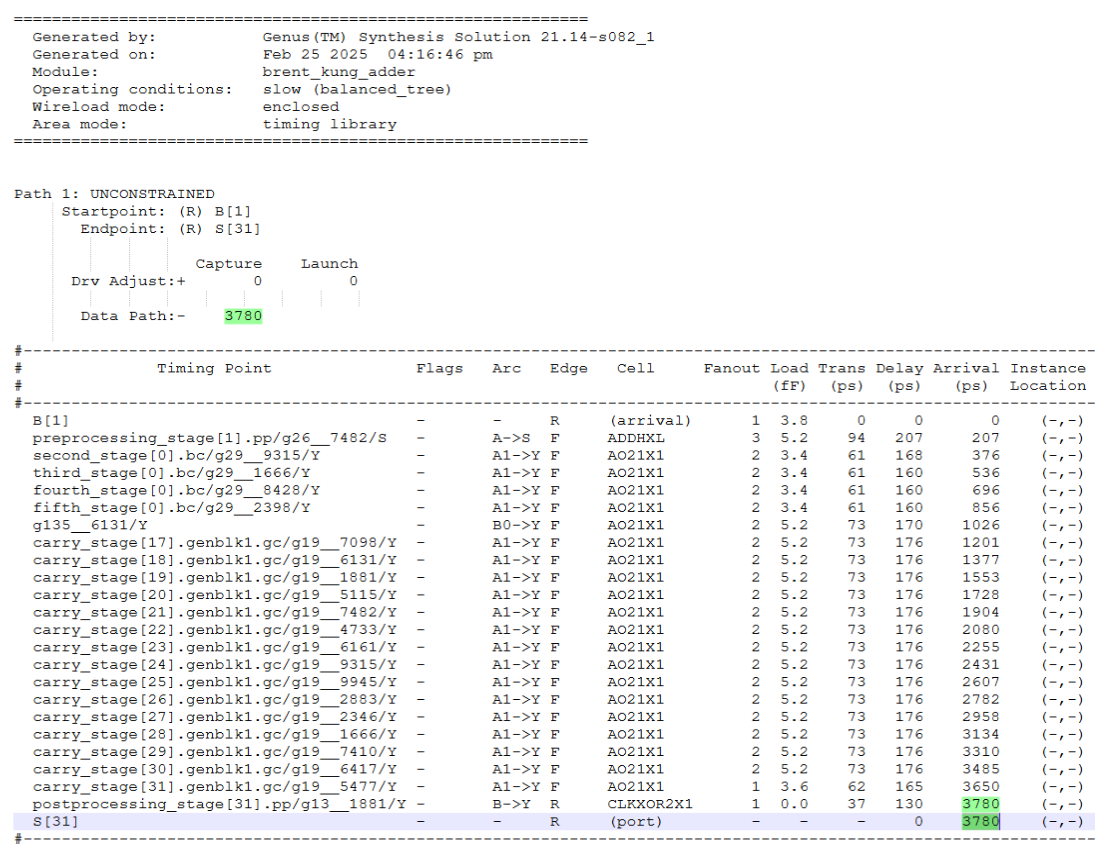}
    \caption{Timing Report}
    \label{fig:enter-label}
\end{figure}
\subsection{Compiled Data of PAT Reports:
}
Data is as follows:\\
Total Power Consumption: \textbf{43.32 $\mu$W}\\
Total Cell Area: \textbf{1223.91 $\mu$m²}\\
Critical Path Delay: \textbf{3.78 ns/3780ps}

\subsection{Comparison with some existing designs:}
To evaluate the effectiveness of the proposed Brent-Kung adder (BKA) architecture, we performed a comparative delay analysis of 32-bit semicustom adder implementations. Table 2 summarizes the critical path delays of eleven adder architectures synthesized under similar semi-custom design methodologies, ensuring a fair comparison of their performance characteristics.

Some Key observations from the analysis include:
\begin{itemize}
    \item \textbf{Delay:} The proposed Brent-Kung Adder achieves a delay of 3.780 ns, outperforming traditional architectures such as the Ripple Carry Adder (RCA, 57.897 ns) and Carry Lookahead Adder (CLA, 44.897 ns). This demonstrates its suitability for high-speed arithmetic applications.
    
    \item \textbf{Parallel Prefix Efficiency:} Parallel prefix adders like the Kogge-Stone Adder (KSA, 21.326 ns) and Spanning Tree Adder (SPA, 31.128 ns) exhibit significantly lower delays than linear architectures (e.g., RCA), validating their logarithmic carry propagation advantage.
\end{itemize}

\noindent \textbf{Full Adder Names in Table:}  
For clarity, the abbreviated adder names in Table \ref{tab:adder_performance} correspond to the following architectures:
Brent-Kung Adder (BKA), Ripple Carry Adder (RCA), Carry Increment Adder (CIA), Carry Lookahead Adder (CLA), Ladner-Fischer Adder (LFA), Kogge-Stone Adder (KSA), Carry Skip Adder (CSKA), Spanning Tree Adder (SPA), Modified Manchester Carry Chain Adder (MMCCA), Sparse Kogge-Stone Adder (SKSA), Han-Carlson Adder (HCA).\\

\begin{table}[h]
    \centering
    \renewcommand{\arraystretch}{1.2} 
    \begin{tabular}{|c|c|c|c|c|}
        \hline
        \textbf{Sr. No.} & \textbf{Adder Type} & \textbf{Delay (ns)} & \textbf{Bit Width} & \textbf{Source} \\ \hline
        1  & BKA   & 3.780  & 32-bit & This Paper \\ 
        2  & RCA   & 57.897 & 32-bit & [4]\\ 
        3  & CIA   & 26.57  & 32-bit & [5] \\ 
        4  & CLA   & 44.897 & 32-bit & [4] \\ 
        5  & LFA   & 21.879 & 32-bit & [6] \\ 
        6  & KSA   & 21.326 & 32-bit & [4] \\ 
        7  & CSKA  & 25.514 & 32-bit & [5]\\ 
        8  & SPA   & 31.128 & 32-bit & [4]\\ 
        9  & MMCCA & 31.87  & 32-bit & [7] \\ 
        10 & SKSA  & 19.895 & 32-bit & [8] \\ 
        11 & HCA   & 0.225  & 32-bit & [9]\\ \hline
    \end{tabular}
    \caption{Adder Performance Comparison (32-bit Semicustom Designs)}
    \label{tab:adder_performance}
\end{table}

\section{Conclusion}
The Brent-Kung Adder (BKA) emerges as an optimal choice for high-speed addition requirements across various digital applications. This 32-bit semicustom frontend design demonstrates superior performance with a critical path delay of 3.78 ns, significantly outperforming traditional architectures like the Ripple Carry Adder (57.897 ns) and the Carry Lookahead Adder (44.897 ns). Additionally, it maintains a balanced power consumption of 43.32 micro-Watts, with internal power contributing 60.09\% of the total. In terms of hardware efficiency, the BKA utilizes an area of 1223.91 $\mu$m², effectively balancing speed and complexity.  

These characteristics make the BKA particularly well-suited for digital VLSI systems requiring fast arithmetic operations, high-performance computing applications, and real-time signal processing tasks. It is especially valuable in scenarios where efficient, high-speed binary addition is critical. While not the absolute fastest among all adder architectures, the BKA offers a compelling trade-off between speed, area, and power consumption. Its logarithmic delay characteristics and optimized carry propagation make it a versatile choice for modern digital system design, particularly in applications where nanosecond-level performance improvements can significantly enhance overall system efficiency.\\
The Codes and Files shown in this paper can be accessed on GitHub[14].

\end{document}